\documentclass[%
 aps,
 amsmath,amssymb,
 reprint,%
 pre,
  normalem,
]{revtex4-2}

\usepackage{graphicx}
\usepackage{dcolumn}
\usepackage{bm}
\usepackage{ulem}
\usepackage[utf8]{inputenc}
\usepackage[T1]{fontenc}

\usepackage{mathptmx}
\usepackage{xcolor}
\usepackage{nomencl}
\makenomenclature

\begin{document}

\preprint{AIP/123-QED}

\title[Coupled electrokinetic transport through a nanoporous membrane]{Coupled electrokinetic transport through a nanoporous membrane: {effects of pore interactions.}}

\author{O. Liot}
\email{olivier.liot@imft.fr}
\affiliation{Universit\'e de Toulouse, INPT, UPS, IMFT (Institut de M\'ecanique des Fluides de Toulouse), All\'ee Camille Soula, F-31400
Toulouse, France}
\affiliation{ 
Institut Lumi\`ere Mati\`ere, Universit\'e Lyon 1, CNRS, UMR 5306, 10 rue Ada Byron, 69622 Villeurbanne Cedex, France
}%
\author{C. Sempere}
\affiliation{ 
Institut Lumi\`ere Mati\`ere, Universit\'e Lyon 1, CNRS, UMR 5306, 10 rue Ada Byron, 69622 Villeurbanne Cedex, France
}%
\author{C. Ybert}%
\affiliation{ 
Institut Lumi\`ere Mati\`ere, Universit\'e Lyon 1, CNRS, UMR 5306, 10 rue Ada Byron, 69622 Villeurbanne Cedex, France 
}%

\author{A.-L. Biance}
\affiliation{ 
Institut Lumi\`ere Mati\`ere, Universit\'e Lyon 1, CNRS, UMR 5306, 10 rue Ada Byron, 69622 Villeurbanne Cedex, France
}%
\date{\today}

\begin{abstract}
{Liquid transport through nanopore is central into many applications, from water purification to biosensing or energy harvesting. Ultimately thin nanopores are of major interest in these applications to increase driving potential and reduce as much as possible dissipation sources. We investigate here the efficiency of the electrical power generation through an ultrathin nanoporous membrane by means of streaming current (electrical current induced by ionic flow in the vicinity of the liquid/solid interface) or electroosmosis (flow rate induced by an electrical potential)}.
{Upscaling from one unique pore to a nanoporous membrane is not straightforward when we consider low aspect ratio nanopore  because of 3D entrance effects, which lead to interactions between the pores. Whereas these interactions have already been considered for direct transport (hydrodynamic permeability of the membrane, ionic conductance), specific effects appear when coupled transports are considered. 
We {obtain} here the expression of the electroosmotic mobility {for a nanoporous membrane} including surface conduction, and {by a numerical evaluation of our results,} we show that (i) it depends mainly of the distance between the pores and (ii) it is sublinear with the number of pores. 
Varying the pore spatial organization (square, hexagonal, disordered structure) reveals that these transport properties are only dependent on one parameter, the porosity of the membrane {(if other parameters such as membrane material/thickness are kept constant)}. Finally, when considering energy conversion yield, it is shown that increasing the number of pores is deleterious, and a non-monotonic behavior with salt concentration is reported.}

\end{abstract}

\maketitle

\section{\label{sec:level1}Introduction}


Transport properties (ionic/chemical/mass/thermal transport) in a liquid at the nanoscale (nanofluidics) is principally affected by surface interactions \cite{schoch2008,bocquet2010,bocquet2014,haywood2015,ghosal2019}. Such transport phenomena{, encountered in natural objects \cite{mclaughlin1981,gravelle2013},} are coupled and have many applications, in biosensing and DNA sequencing \cite{rems2016}, in {material science \cite{xu2018, bonhomme2020}}, and, more importantly, in alb{green} energy harvesting \cite{wang2017} (thermal or chemical energy in electricity for example \cite{van_der_heyden2006, siria2013,sempere2015,feng2016}), {for} water treatment \cite{malaeb2011} or desalination \cite{achilli2010,cohen-tanugi2012}. As another example, pierced graphene monolayer has been the ultimate target for water filtration \cite{qi2018}, which involves simultaneously fluid transport through a membrane and electrostatic exclusion by the pores.

{Coupled transports (eletrokinetic phenomena -- EK) are defined as transports of a quantity which is not driven by its ``natural'' forcing. They result from interactions between bulk and surfaces properties. Use of nanofluidic-devices such as membrane pierced of nanopores or nanochannels, wherein the surface-to-volume ratio is large, favour these kinds of transports. In an insulating solid nanochannel containing an ionic solution, some ionic charges are present at the liquid-solid interface and a diffuse layer of counter ions (electrical double layer, EDL) forms in the vicinity of the surface. The typical width of this layer is the Debye length and scales as \cite{schoch2008}: 
\begin{equation}
    \kappa^{-1}\propto\frac{1}{\sqrt{I_s}},
\end{equation}
\noindent where $I_s=\sum_i c_iz_i^2$ is the ionic strength of the solution, with $c_i$ the concentration of ionic specie $i$ and $z_i$ its valence.  We can cite four of them: (i) electroosmosis is the flow generated by a difference of electric potential; (ii) streaming current is the ionic transport driven by a pressure drop; (iii) thermoosmosis is the fluid flow generated by a temperature difference; (iv) diffusioosmosis, provoked by salt gradient, is characterized both by fluid and ion transports.}

In this context, the use of nanopores (ultrathin channels with low aspect ratio) 
seems promising to increase potential gradient (inversely proportional to the channel length), reduce dissipation and then enhance EK transport and energy conversion yield. In this nanopore geometry, however, so-called entrance effects, due to convergence of flow streams at the entrance of the pore,  have to be taken into account. Entrance hydrodynamic permeability  has been derived for more than one century \cite{couette1890}, whereas entrance effects concerning ionic \cite{lee2012}, electroosmostic \cite{mao2014} and diffusio-osmotic transport \cite{rankin2019} is still a subject of active research.



{However, }most of studies so far focus on one nanopore or one nanochannel. 
Nevertheless, to design macroscopic membranes with large pore density for energy harvesting, some insights are compulsory to understand the influence of the interactions between pores during EK transport. 
Sub-additive electric conductance was experimentally observed and modelled for ionic transport through an assembly of nanopores \cite{gadaleta2014}. It leads to a pore conductance scaling as the square root of the number of pores for a 2D array {made of parallel pores}. In addition, the more the pore are isolated, the higher will be the resulting current \cite{green2015-1}. {Despite a seemingly analogous system, when considering the hydrodynamic permeability, theoretical analysis \cite{jensen2014} and experiments \cite{sempere2015} show that pore interactions result in a slight increase of the permeability of the membrane.

However, to our knowledge, no study has focused yet on pore interaction influence on coupled transport properties and especially on energy harvesting efficiency, although it} is central to understand the salient features necessary to improve and optimise applied systems such as osmotic energy power plants \cite{van_der_heyden2006,sempere2015}. This work aims to study both analytically and numerically the coupled electrokinetic transports (in particular streaming current and electroosmosis) through a multi-pore membrane. In particular, the effects of the spatial arrangements of the pores (ordered and disordered membranes) and of the number of pores on energy harvesting efficiency and yield is investigated.

{The article is organized as follows. {First, we determine} the electrokinetic transport coefficients for one pore taking into account entrance effects and, then, for an assembly of pore. Secondly, {we numerically evaluate our results of the first part, for a membrane with pores of given characteristics. In particular, we show how the number of pores and their spatial arrangement on the membrane modify the electrokinetic response}. Finally, energy harvesting yield is calculated for different cases and analyzed.}

\section{Determination of the electrokinetic transport coefficients}
Compared to previous works on electric \cite{gadaleta2014} and fluidic \cite{jensen2014,sempere2015} transport through multi-pore membrane, the main difficulty for studying streaming current or electroosmosis lies in the coupling between ions and fluid transports. {These couplings don't depend only on inner transport properties but also entrance effects. These entrance effects are modified by pore interactions, so it is necessary to investigate the relationships between couplings, entrance effects and pore interactions.} Moreover, surface-conduction dominated electric transport through nanopores \cite{lee2012} could affect coupled transport too. {We will first {recall} the transport properties of one nanopore, taking into account entrance effects. We will then consider an assembly of nanopores and {compute} scaling laws to predict how the transport is affected by interactions with neighbouring pores. Finally, we will {deduce} the full expression of EK transport coefficients for a membrane with $N$ pores organized with a known geometry, taking into account entrance effects and surface conduction.} 

\subsection{Electrokinetic transport coefficients for one pore}

{T}he electrokinetic coupling for one pore is given by the linear response theory, formalized here by the Onsager matrix \cite{onsager1931}: 

\begin{equation}
    \begin{pmatrix}
Q \\
I
\end{pmatrix}=
\begin{pmatrix}
K_h & \mu_{eo} \\
\mu_{eo} & K_e
\end{pmatrix}
\begin{pmatrix}
\Delta P\\
\Delta V
\end{pmatrix}.
\label{eq:onsager1}
\end{equation}

\noindent We denote $Q$ the fluid flow rate through the pore, $I$ the electrical current, $\Delta P$ the external pressure gradient and $\Delta V$ the potential difference applied to the pore. The coefficients of the Onsager matrix are: $K_h$ the hydrodynamic permeability of the pore, $K_e$ the electric conductance, and $\mu_{eo}$ the electroosmotic mobility. Note that due to Onsager reciprocity, the same coefficient $\mu_{eo}$ appears for both electroosmosis and streaming current.

\begin{figure}[h!]
    \centering
    \includegraphics[width=0.95\linewidth]{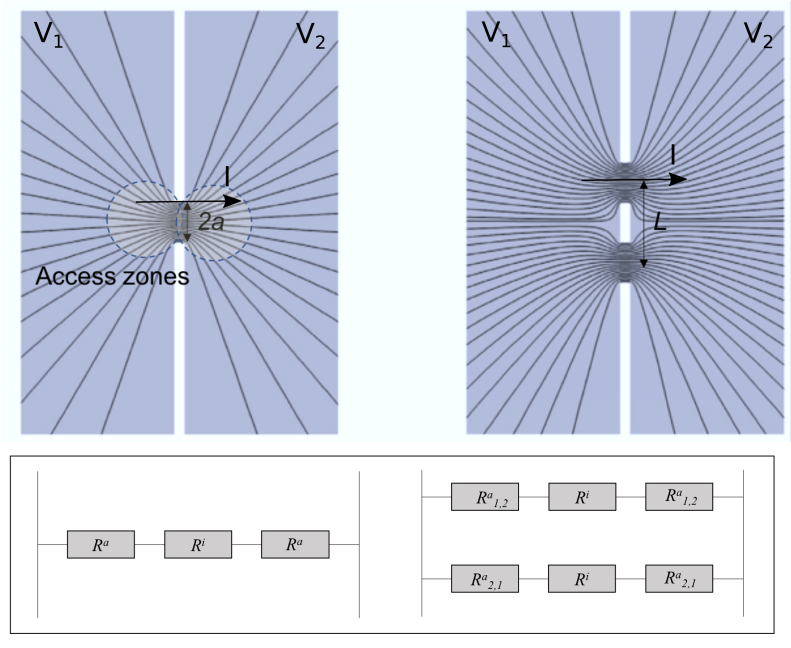}
    \caption{{Top:} Electric field lines converging towards a single pore (left) or two pores (right). {Driving force ($\Delta V=V_1-V_2$) and flux ($I$) are mentioned. {Bottom: Equivalent electrical (or hydrodynamic) circuit showing for one (left) and two (right) pores. The access and inner resistances are in series for each pore, and the pores are in parallel. Access resistances are modified by the environment, and in particular by the presence of other pores.}}}
    \label{scheme}
\end{figure}

We decompose the problem in two parts : the inner part inside the nanopore and the access part, as depicted in Fig.~\ref{scheme}. 
The fluxes (flow rate, ionic current) are conserved in the different zones because access regions and pore are serial connected, {then can be simply added in the linear regime, as shown theoretically and experimentally for hydrodynamic and ionic transport \cite{lee2012, gadaleta2015}.} {Fig. \ref{scheme}, bottom, shows the analog circuit considered here}.
These flux can then be linked to local quantities applying in these zones. For the inner pore region, it reads:

\begin{equation}
    \begin{pmatrix}
Q \\
I
\end{pmatrix}=
\begin{pmatrix}
K_h^i & \mu_{eo}^i \\
\mu_{eo}^i & K_e^i
\end{pmatrix}
\begin{pmatrix}
\Delta P^i\\
\Delta V^i
\end{pmatrix},
\end{equation}

\noindent where index $i$ is for the inside-pore region. {This expression is valid in the limit of low currents, which is the scope of our study.} The transport coefficient inside the pore ($K_h^i$, $\mu_{eo}^i$ and $K_e^i$) are well-known. 

Concerning the access region, indexed by $a$, we neglect cross phenomena {that} could occur in the diffusive layer of the membrane \cite{mao2013}, which means that there is no electroosmotic flow outside the pore. 
Thus, fluxes are similarly connected to local properties: 

\begin{equation}
    \begin{pmatrix}
Q \\
I
\end{pmatrix}=
\begin{pmatrix}
K_h^a & 0 \\
0 & K_e^a
\end{pmatrix}
\begin{pmatrix}
\Delta P^a\\
\Delta V^a
\end{pmatrix}.
\end{equation}
{Note that the relative contribution of access and inner regions depend on the aspect ratio of the pore \cite{sampson1891}, but also on other parameters characterizing interfacial properties (hydrodynamic friction \cite{gravelle2013} or surface charge density \cite{lee2012} for example).}
Considering that the nanopore is symmetric, two access contributions shall be considered, the pore's inlet and outlet. Then, the potentials can be decomposed as:
\begin{eqnarray}
\Delta P&=&\Delta P^i+2\Delta P^a,\label{eq:decompoDP}\\
\Delta V &=& \Delta V^i+2\Delta V^a.\label{eq:decompoDV}
\end{eqnarray}
 
\noindent {Donnan potential \cite{schoch2008} is neglected as we consider the linear low current limit, low voltage and non-overlapping EDL.} {To determine electrical conductivity and electroosmotic mobility of the system}, we first assume a configuration where no pressure gradient is applied ($\Delta P=0$). Balancing fluxes in access and inner regions {we get:}

\begin{eqnarray}
    Q&=&\mu_{eo}\Delta V=K_h^i\Delta P^i+\mu_{eo}^i\Delta V^i=K_h^a\Delta P^a,\label{eq:surQ}\\
    I&=&K_e\Delta V=\mu_{eo}^i\Delta P^i+K_e^i\Delta V^i=K_e^a\Delta V^a.\label{eq:surI}
\end{eqnarray}

\noindent {Using eqs. \ref{eq:decompoDP} and \ref{eq:decompoDV}, and the fact that $\Delta P=0$ we can write:}

\begin{eqnarray}
   {\mu_{eo}\Delta V}&=&{-K_h^a\dfrac{\Delta P^i}{2},}\\
   {K_e\Delta V}&=&{K_e^a\dfrac{\Delta V-\Delta V^i}{2}.}
\end{eqnarray}

\noindent It leads to the following relations between the potentials: 

\begin{eqnarray}
    \Delta P^i&=&-2\frac{\mu_{eo}}{K_h^a}\Delta V\\
    \Delta V^i&=&\left(1-2\frac{K_e}{K_e^a}\right)\Delta V.
\end{eqnarray}

Inserting these in {the inner-transport-related term of} eqs. \ref{eq:surQ} and \ref{eq:surI}, we get a linear system on $K_e$ and $\mu_{eo}$:

\begin{eqnarray}
    {\left(1+2\dfrac{K_h^i}{K_h^a}+\right)\mu_{eo}+2\dfrac{\mu_{eo}^i}{K_e^a}K_e}&=&{\mu_{eo}^i,}\\
    {2\dfrac{\mu_{eo}^i}{K_h^a}\mu_{eo}+\left(1+2\dfrac{K_h^i}{K_h^a}\right)K_e}&=&{K_e^i.}
\end{eqnarray}

\noindent {It} can be solved as:

\begin{eqnarray}
    K_e&=&\frac{1+2k_h(1-\alpha^i)}{1+2(k_e+k_h)+4k_ek_h(1-\alpha^i)}K_e^i, \label{eq:Ke}\\
    \mu_{eo}&=&\frac{1}{1+2(k_e+k_h)+4k_ek_h(1-\alpha^i)}\mu_{eo}^i. \label{eq:mueo}
\end{eqnarray}

\noindent with $k_h=K_h^i/K_h^a$, $k_e=K_e^i/K_e^a$ and $\alpha^i=(\mu_{eo}^i)^2/(K_e^iK_h^i)$. To get the hydrodynamic permeability, we follow the same reasoning in a configuration where $\Delta V=0$, which results in 

\begin{equation}
    K_h=\frac{1+2k_e(1-\alpha^i)}{1+2(k_e+k_h)+4k_ek_h(1-\alpha^i)}K_h^i,
    \label{eq:Kh}
\end{equation}

\noindent and we obtain the same expression as previously for $\mu_{eo}$, {satisfying Onsager reciprocity theorem \cite{onsager1931}}. {Eqs. \ref{eq:Ke}, \ref{eq:mueo} and \ref{eq:Kh} link the linear response coefficients of one pore, taking into account both inner and access regions. These relationships are general (for symmetric pores), but the precise coefficients need to be explicitly computed for each situation, as they depend on many factors such as the size of the systems, the surface properties and the pore environment as detailed in the following.}

\subsection{Electroosmotic mobility for a multi-pore membrane: effect of the number of pores.}

We now consider a 2D array of $N$  pores of radius $a$ on a membrane of thickness $h$ with $h=2a$. {We fix this thickness to allow future experimental comparisons with pierced silicon nitride membranes used by Gadaletta \emph{et al.} \cite{gadaleta2014}.} 
We focus {on determining how }the total coupled transport coefficient  depends on the number of pores crossing the membrane. 
On the one hand, coefficients corresponding to inner do not include entrance effects and so pore interactions. Inner hydraulic and electric resistances are independent, and nanopores are disposed in parallel from the hydrodynamic and electrical point of view. Then, when $N \rightarrow \infty$, $K_{e,N}^i\propto N K_{e}^i$, $K_{h,N}^i\propto N K_{h}^i$ and $\mu_{eo,N}^i\propto N \mu_{eo}^i$. The subscript $N$ refers to the total value for $N$ pores.

For the access quantity, previous work \cite{gadaleta2014} showed that $K_{e,N}^a\propto \sqrt{N}$. Concerning the hydrodynamic transport, the physical origin of an entrance effect is the focusing of streamlines at the interface between large reservoir and thin pore. Since a viscous flow dissipates energy when streamlines change in direction, we have an access resistance $R_h^a=1/K_h^a$. Previous works \cite{couette1890,sampson1891,happel1983} showed that $R_h^a=\frac{3\eta}{2a^3}$ where $\eta$ is the dynamic viscosity of the fluid. {This access resistance is the same for fluid entering inside the pore and for fluid leaving it.} Other studies \cite{weissberg1962,dagan1982} put forward that the total resistance of the pore crossing a membrane is very close to the sum of Poiseuille and access resistances. In the case of membranes where $h\sim a$, the access resistance cannot be neglected. Indeed $R_h^i=1/K_h^i=8\eta h/(\pi a^4)$ so $R^a_h/R_h^i\sim h/a\sim 1$.

To take into account hydrodynamic interactions between pores, we use previous work from Tio and Sadhal \cite{tio1994}. {At first order, only pairwise interactions are considered.} They showed that the access resistance of a pore labelled $m$ among other pores labeled $n$ is:

\begin{eqnarray}
    R^{a,(m)}_h&=&\frac{3\eta}{2a^3}\left(1-\sum_{n,n\neq m}\left[\frac{2}{3\pi}\left(\frac{a}{L_{n,m}}\right)^3+\frac{6}{5\pi}\left(\frac{a}{L_{n,m}}\right)^5\right.\right.\nonumber\\
    &&+\left.\left.\frac{18}{7\pi}\left(\frac{a}{L_{n,m}}\right)^7+...\right]\right). \label{eq:Rham}
\end{eqnarray}

\noindent $L_{n,m}$ represents the center-to-center distance between pore $m$ and pore $n$. We can shrink the previous equation, introducing $\lambda^{(m)}$:

\begin{equation}
        R^{a,(m)}_h=R^{a}_h(1-\lambda^{(m)}).
\end{equation}


{We denote as $L$ the typical distance between two pore centers. When $L/a>3$, i.e. space between two pores' edge is {larger than} one pore radius, we can consider only the first term of $\lambda^{(m)}$. For $N\rightarrow\infty$, $\sum_{n,n\neq m}\left(\frac{1}{|r_{n,m}|}\right)^3$ tends towards a constant ({it can be found numerically that this constant is }close to 11 for an hexagonal array). Thus, since $\lambda^{(m)}$ tends to a constant $\Lambda$ for large $N$, all the pores have the same access hydrodynamic resistance. These resistances can finally be considered in parallel, as described in fig. \ref{scheme},} {as they have the same pressure drop through. Thus}, we get :
\begin{equation}
    K_{h,N}^a=\frac{1}{R_{h,N}^a}=\sum_m \frac{1}{R^{a,(m)}_h}=K_h^a\sum_m\frac{1}{1-\Lambda}.
\end{equation}

\noindent We {finally} have the following scaling:

\begin{equation}
    K_{h,N}^a\approx K_h^aN\frac{1}{1-\Lambda}\propto N.
\end{equation}

{Despite a seemingly analogous problem, we observe a large difference with electrical conductance. {Whereas pore interactions tend to a sub-linear behaviour of $K_{e,N}^a$, we obtain a linear scaling for the access hydrodynamic permeability $K_{h,N}^a$.}  Due to the constant correction $1/(1-\Lambda)>1$, we have here a flow rate increase thanks to hydrodynamic interactions between pores, as observed previously \cite{jensen2014}.}

{We search now for a scaling of $\mu_{eo,N}$. We inject in eq. \ref{eq:mueo} the scalings obtained for the access electric conductance ($K^a_{e,N}\propto\sqrt{N}$) and the access hydrodynamic permeability ($K^a_{h,N}\propto N$). After simplifications, it reads for large $N$:}


\begin{equation}
    \mu_{eo,N}\propto\sqrt{N}.                 
\end{equation}

Similarly to the electric transport \cite{gadaleta2014}, the electroosmotic transport adopts a sub-linear scaling with the number of pores.

\subsection{Transport coefficients for a membrane pierced with N identical pores with {any} spatial organization.}

To upscale experimentally the coupled transport value through a large multi-pore membrane, some studies {at the scale of a few pores have allowed} to capture main interaction mechanisms between the pores. Such an approach has been used to investigate filtration by dedicated membrane for example \cite{liot2018,van_zwieten2018,sauret2018} or for the study of electric transport (direct transport) \cite{gadaleta2014,green2015-1}. Numerically, we are able to by-pass this step and estimate directly the behaviour of a very large membrane. {{In addition to pore geometry, membrane material and thickness, geometry of the reservoirs, salt type and concentration, p}revious works \cite{jensen2014, gadaleta2014}} {suggest} that two main parameters drive the membrane response to an external driving: inter-pore distance and lattice organization. We develop here a model for a membrane pierced by $N$ pores of the same size and organized in a known geometry (such as hexagonal lattice, square lattice or randomly-ordered pores) and we show its effect on transport coefficients. 
{We then want to specify the coefficients $K_e$, $K_h$ and $\mu_{eo}$ obtained in the previous subsection, eqs. \ref{eq:Ke}, \ref{eq:mueo} and \ref{eq:Kh}}. 

We recall first the expressions of the inner Onsager coefficients for a cylindrical nanopore, {in the limit of small zeta potentials and no inertia} \cite{happel1983,lee2012,mao2014} : 

\begin{eqnarray}
K^i_h&=&\frac{\pi a^4}{8\eta h},\\
K^i_e&=&\kappa_b\frac{\pi a^2}{h}+\kappa_s\frac{2\pi a}{h},\\
\mu_{eo}^i&=&-\frac{\epsilon_0 \epsilon_r \zeta}{\eta}\frac{\pi a^2}{h}.
\label{eq:innercoef}
\end{eqnarray}

\noindent with $\eta$ the fluid dynamic viscosity, $\epsilon_0 \epsilon_r$ is the fluid dielectric constant, $\kappa_b$ the bulk {conductivity} of the solution and $\zeta$ the membrane zeta potential. {These expressions of $K^i_e$ and $\mu_{eo}^i$} {are taken in the limit of no EDL overlap, $a>\kappa^{-1}$.} {These expressions of electric conductance and electroosmotic mobility are valid in the limit $h>a$, \cite{yariv2015}.} The electric conductance takes into account surface effects inside the pore, with $\kappa_s=\kappa_b\frac{|\Sigma|}{2ec_0}$ the surface conductivity \cite{lee2012}. $\Sigma$ is the surface charge density, $e$ the elementary electric charge and $c_0$ the electrolyte concentration. We denote as $l_{Du}=\kappa_s/\kappa_b$ the Dukhin length which represents the competition between bulk and surface conduction \cite{bocquet2010}.

The access direct transport coefficients have also been derived \cite{happel1983} {and numerically and experimentally determined \cite{lee2012}}:

\begin{eqnarray}
K^a_h&=&\frac{2a^3}{3\eta},\\
K^a_e&=&2\kappa_b\left(2a+l_{Du}\right). \label{eq:kae}
\end{eqnarray}

\noindent Note that the access electric conductance also takes into account {a surface contribution} proposed by Lee {et al.} \cite{lee2012}.

From these results we can explicit the following coefficients: 

\begin{eqnarray}
k_h&=&\frac{3\pi a}{16h},\label{eq:kh}\\
k_e&=&\frac{\pi a}{2h}\left(\frac{a+2l_{Du}}{2a+l_{Du}}\right),\label{eq:ke}\\
\alpha^i&=&\frac{8(\epsilon_0 \epsilon_r)^2\zeta^2}{\eta a\kappa_b\left(a+2l_{Du}\right)}.
\end{eqnarray}

To keep only dominant terms in expressions of $K_e$, $K_h$ and $\mu_{eo}$, we estimate the order of magnitude of $\alpha^i$ for common situations. We consider water with dissolved $K^+$ and $Cl^-$ ions, flowing through a membrane of silicon nitride (Si$_3$N$_4$) at ambient temperature and $pH\approx 6$, values consistent with previous experimental works \cite{lee2012,gadaleta2014}. Table \ref{table:1} lists the values used to estimate $\alpha^i$. We recall that $\kappa_b=(\lambda_{K^+}+\lambda_{Cl^-})c_0$ where $\lambda_{K^+}$ and $\lambda_{Cl^-}$ are the molar ionic conductivity for $K^+$ and $Cl^-$ respectively. We consider concentrations between 0.001\,mol.L$^{-1}$ (M) and 1\,mol.L$^{-1}$. In addition, we set $|\Sigma|=20\,$mC.m$^{-2}$ \cite{lee2012}.

\begin{table}[h!]
\begin{tabular}{c|c|c|c}
    Name & Symbol & Value & Unit\\
    \hline
    \hline
   Dielectric constant & $\epsilon_0\epsilon_r$ & $6.95\times 10^{-10}$ & F.m$^{-1}$  \\
   Zeta potential & $\zeta$ & -30 & mV \cite{bousse1991}\\
   Dynamic viscosity & $\eta$ & $1.0\times10^{-3}$ & Pa.s\\
   Pore radius & $a$ & 25 & nm\\
   Salt concentration & $c_0$ & $0.001 - 1$ & mol.L$^{-1}$\\
   Bulk conductivity & $\kappa_b$ & $1.49\times 10^{-2}-14.9$& S.m$^{-1}$\\
   Dukhin length & $l_{Du}$ &$0.104-104$ & nm\\
   K$^+$ ionic conductivity & $\lambda_{K^+}$ &$7.35$ & mS.m$^2$.mol$^{-1}$\\
   Cl$^-$ ionic conductivity & $\lambda_{Cl^-}$ &$7.63$ & mS.m$^2$.mol$^{-1}$\\
   \hline
\end{tabular}
\caption{Values used to compute $\alpha^i$.}
\label{table:1}
\end{table}

Within these conditions, $\alpha^i$ varies between $1.2\times10^{-4}$ and $2.1\times10^{-2}$, thus $\alpha^i\ll 1$. We then approximate the electric conductance, hydraulic permeability and electroosmotic mobility {obtained at eqs. \ref{eq:Ke}, \ref{eq:Kh} and \ref{eq:mueo}} by: 

\begin{eqnarray}
    K_h&\simeq&\frac{1}{1+2k_h}K_h^i,\\
    K_e&\simeq&\frac{1}{1+2k_e}K_e^i,\\
    \mu_{eo}&\simeq&\frac{1}{(1+2\,k_e)(1+2\,k_h)}\mu_{eo}^i. 
\end{eqnarray}

\noindent We observe that direct transport coefficients, $K_h$ and $K_e$, do not include coupling term: electric conductance depends only on electric terms and hydraulic conductance is a function of hydraulic quantities only. Coupling effects affect only the electroosmotic mobility. We simplify these expressions by defining the following coefficients: 

\begin{eqnarray}
    \beta_h&=&\frac{1}{1+2k_h},\\
    \beta_e&=&\frac{1}{1+2k_e},\\
    \beta_{eo}&=&\frac{1}{(1+2\,k_e)(1+2\,k_h)}.
\end{eqnarray}

\noindent We notice that $\beta_{eo}=\beta_e\beta_h$. As expected, these coefficients are lower than 1, {because access resistance has indeed to be added to the inner one}. 

Actually, these coefficients $\beta$ are relative to the one-pore case. They include implicitly, {via} $k_e$ and $k_h$, the access hydraulic permeability and electric conductance. For a multi-pore alb{membrane}, each $\beta$ is different for each pore and depends on the spatial environment made by the other pores. Following the same notations as in the previous subsection, we name $\beta_h^{(m)}$, $\beta_e^{(m)}$ and $\beta_{eo}^{(m)}$ the coefficients $\beta$ for the pore $m$ for the three kinds of transport (hydraulic, electric, coupled). So for each pore $m$ we { define $\beta_h
^{(m)}$ such as} $K_h^{(m)}=\beta_h^{(m)}K_h^i$ (note that inner coefficients are independent of the considered pore, if we consider identical pores). We have the analogous relations for the electric and coupled transports. Using the same convention, {we define:} 

\begin{eqnarray}
    \beta_h^{(m)}&=&\frac{1}{1+2k_h^{(m)}},\\
    \beta_e^{(m)}&=&\frac{1}{1+2k_e^{(m)}},\\
    \beta_{eo}^{(m)}&=&\frac{1}{(1+2\,k_e^{(m)})(1+2\,k_h^{(m)})},
\end{eqnarray}

\noindent with $k_h^{(m)}=K_h^i/K^{a,(m)}_h$ and $k_e^{(m)}=K_e^i/K^{a,(m)}_e$. {One notices again that $\beta_{eo}^{(m)}=\beta_e^{(m)}\beta_h^{(m)}$. In previous works \cite{gadaleta2014,jensen2014}, it has been shown numerically and experimentally that $1-\beta_e^{(m)} \gg 1-\beta_h^{(m)}$, so at first sight, $\beta_{eo}^{(m)} \sim \beta_e^{(m)}$, meaning that the effect of pore interactions on the electroosmotic response is very similar to the effect of the electric response.}

$K_h^{a,(m)}$ can be determined by reversing eq.\ref{eq:Rham}. {The estimation of $K_e^{a,(m)}$ and $\mu_{eo}^{a,(m)}$ necessitates} the estimation of $\beta_{eo}^{(m)}$. 
{We need to extend the work from Gadaleta {et al.} \cite{gadaleta2014} to consider surface conduction. They consider pairwise interactions between access zones of each pore and deduce the total electric conductance of the membrane, including the access one. Practically, they use an analogy between ion transport and electrostatic capacitance. The pore is considered as a infinitely thin disk which holds an electric charge. This charge is related to the disk (pore) radius, and is influenced by neighbouring pores. Nevertheless, they neglect surface conduction ($l_{Du}\ll a$). Lee et al. \cite{lee2012} showed that due to surface conduction, the apparent radius of the access zone for electrical conductance is $a^{eff}=a+0.5\,l_{Du}$. The ``charge'' held by the pore entrance, as described by Gadaleta et al. \cite{gadaleta2014}, is now held by a disk of radius $a^{eff}$. Thus, combining works from Gadaleta et al. and Lee et al., w}e write (introducing $\gamma^{(m)}$ for the right-hand term with the sum): 


\begin{eqnarray}
    K^{a,(m)}_e&=&2\kappa_b(2a+l_{Du})\left(1+\sum_{n,n\neq m}\frac{a+0.5\,l_{Du}}{L_{n,m}}\right)^{-1} \label{eq:accessKam}\\
    K^{a,(m)}_e&=&2\kappa_b(2a+l_{Du})\left(1+\gamma^{(m)}\right)^{-1}.
\end{eqnarray}

\noindent and we obtain the following equations for $k_e^{(m)}$ and $k_h^{(m)}$:

\begin{eqnarray}
    k_h^{(m)}&=&\frac{3\pi a}{16h}\left(1-\lambda^{(m)}\right)=k_h\left(1-\lambda^{(m)}\right),\\
    k_e^{(m)}&=&\frac{\pi a}{2h}\left(\frac{a+2l_{Du}}{2a+l_{Du}}\right)(1+\gamma^{(m)})= k_e(1+\gamma^{(m)}).
\end{eqnarray}

\noindent One can notice that for an isolated pore, $\gamma^{(m)}=\lambda^{(m)}=0$ and we {recover} expressions of eqs. \ref{eq:kh} and \ref{eq:ke}. 


The total hydraulic permeability, electric conductance and electroosmotic mobility can be computed for membrane pierced with $N$ pores, provided we know all the distances between each pores in the membrane. The coefficients read

\begin{eqnarray}
   K_{h,N}&=&K_h^i\sum_m \beta_h^{(m)},\\
   K_{e,N}&=&K_e^i\sum_m \beta_e^{(m)},\\
   \mu_{eo,N}&=&\mu_{eo}^i\sum_m\beta_{eo}^{(m)}.
\end{eqnarray}

To {determine the effect of interactions between pores, we use the dimensionless expressions for the hydraulic permeability, electric conductance and electroosmotic mobility, normalized by the case of N isolated and parallel pores}:

\begin{eqnarray}
   K_{h,N}^{isolated}&=&NK_h=N\beta_h K_h^i,\\
   K_{e,N}^{isolated}&=&NK_e=N\beta_e K_e^i,\\
   \mu_{eo,N}^{isolated}&=&N\mu_{eo}=N\beta_{eo} \mu_{eo}^i.
\end{eqnarray}

Combining above expressions, we can write as an example the ratio between the total access hydraulic permeability and the one for N isolated pores:
\begin{equation}
    \frac{K_{h,N}}{K_{h,N}^{isolated}}=\frac{\sum_m\beta_h^{(m)}}{N\,\beta_h}.
\end{equation}
\section{Numerical {evaluations} for different types of membranes}

{As the results obtained are not explicit, we propose a numerical evaluation of our previous developments} for large membranes and three different types of pore organizations: hexagonal, square and random. To make the computations for large $N$ ($N\sim10^7$) easier, all pores are considered in the same environment relatively to other pores, {meaning that if we consider a plane surface, pores at the edge are neglected. However, this situation corresponds to pores pierced on a spherical membrane {with the limit of sphere radius being large compared to pore dimensions, to be able to consider the interior of the sphere as a reservoir. This} situation {is} commonly encountered} in natural objects (electrokinetic transport in cells \cite{mclaughlin1981}), bio-assays \cite{vasilca2018} or industrial devices (bed reactors \cite{hartig1993}). For a given $N$, two parameters are relevant for each membrane: the ratio between pore radius and inter-pore distance; the ratio between pore radius and Dukhin length (directly related to the salt concentration if the surface charge is kept constant).


\subsection{Hydrodynamic transport}
\begin{figure}[h!]
    \centering
    \includegraphics[width=0.95\linewidth]{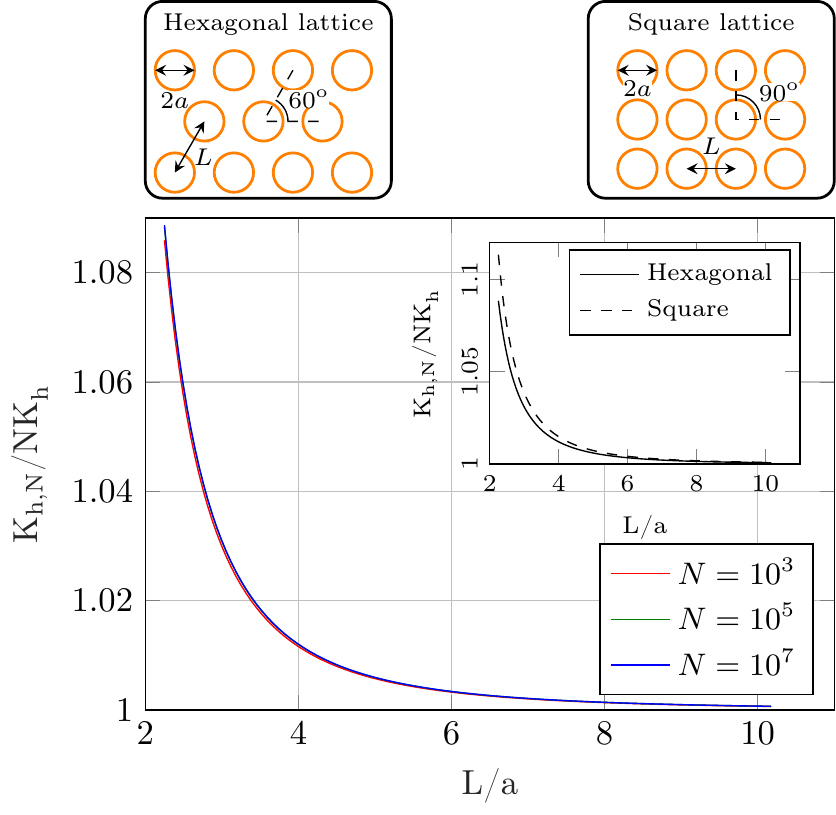}
    \caption{Ratio between total hydraulic permeability $K_{h,N}$ and the one for N isolated pore $K_{h,N}^{isolated}=NK_h$ versus $L/a$ for a square-lattice membrane. Three $N$ are plotted. For the computation, we used $h=50\,$nm and $a=25\,$nm. {Inset: Same quantitiy but for different pore arrangements, $N=10^5$. }{Hexagonal and square lattices are represented above the graph.}}
    \label{KhNsurNKh_square}
\end{figure}

We consider a square-lattice membrane. Fig. \ref{KhNsurNKh_square} shows $K_{h,N} / NK_{h}$ as a function of $L/a$ for {various} $N$. To compute it for an aqueous solution of potassium chloride, we set $a=25\,$nm and $h=50\,$nm. We first observe that for $L/a>10$, the ratio tends asymptotically to 1, which is consistent to more and more isolated pores (and so to decreasing pore interactions){, as previously observed \cite{jensen2014}}.  For small $L/a$, the hydraulic permeability is higher than the ones for isolated pores. The interactions {between the pores} tend to increase the hydrodynamic transport efficiency if pores are close enough to each others, {as observed previously}. We do not observe {any} influence of the number of pores since the hydrodynamic pore interaction is a short-range effect scaling as $(L/a)^{-3}$ at best. Finally, the expressions of $\beta_h$ and $\beta_h^{(m)}$ do not imply electric component so the hydrodynamic transport does not depend on the electrolyte concentration.

\begin{figure}[h!]
    \centering
    \includegraphics[width=0.95\linewidth]{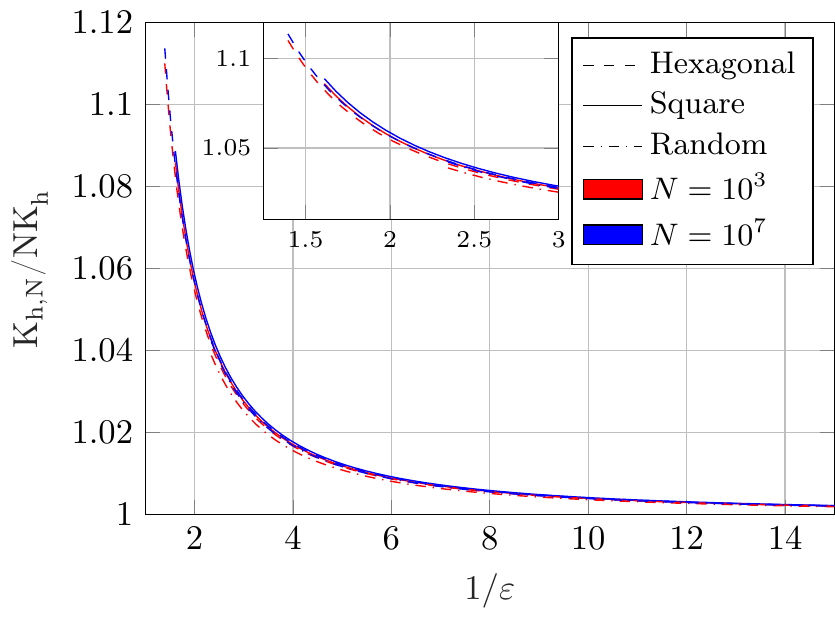}
    \caption{Ratio between total hydraulic permeability $K_{h,N}$ and the one for $N$ isolated pore $K_{h,N}^{isolated}=NK_h$ versus $1/\varepsilon$. Square lattice, hexagonal lattice and random lattice are compared. Two $N$ are plotted. Inset is a zoom of the main plot. For the computation, we used $h=50\,$nm, $a=25\,$nm.}
    \label{KhNsurNKh_comparaison}
\end{figure}

Fig.~\ref{KhNsurNKh_comparaison} shows a comparison of $\frac{K_{h,N}}{NK_{h}}$ between three different membrane configurations: hexagonal lattice, square lattice and random lattice. The last one corresponds to 2D Random Sequential Adsorption \cite{wang1994} whose maximal porosity $\varepsilon$ is about 0.46. {The} porosity {$\epsilon$} is defined as the ratio between {the surfaces of the pore to the total surface of the membrane}:
{
\begin{equation}
\epsilon=\frac{\Sigma_m \pi a_{m}^2}{S_{membrane}}.
\end{equation}
}

\noindent{where $a_m$ is the radius of pore $m$ and $S_{membrane}$ the membrane area.} Fig.~\ref{KhNsurNKh_comparaison} shows curves plotted as a function of $1/\varepsilon$ instead of $L/a$ to normalize the different configurations but keeping the same trend: when $1/\varepsilon$ rises, $L/a$ too. {On the contrary to previous results {(Jensen \emph{et al.}  \cite{jensen2014} claim a difference of hydrodynamic permeability between hexagonal-lattice and square-lattice membranes)} {and to data of fig. \ref{KhNsurNKh_square} (inset)}, we do not observe any noticeable differences when changing the pore spatial organisation. Jensen {et al.} \cite{jensen2014} made computations using only the third order in $a/L_{n,m}$ (see eq. \ref{eq:Rham}) whereas we took as far as the 9\textsuperscript{th} order{, but we did not observe significant difference due to the order choice. Actually} the difference observed is due to changes in porosity. For a given $L/a$, membrane porosity depends on the pore spatial arrangement. Since the porosity is higher for a hexagonal lattice than for a square lattice, it is unavoidable to have lower hydraulic resistance. } The porosity is {actually} the only relevant parameter {and normalisation using $\varepsilon$ collapses the curves of hydraulic permeability}. Inset in fig.~\ref{KhNsurNKh_comparaison} shows a zoom at small $1/\varepsilon$. It reveals very slight differences at high porosity, but the highest accessible porosity is different for each configuration: $\frac{K_{h,N}}{NK_{h}}$ ratio can reach higher values for hexagonal lattice which can access a higher porosity.

\subsection{Electrical transport}

\begin{figure}[h!]
    \centering
    \includegraphics[width=0.95\linewidth]{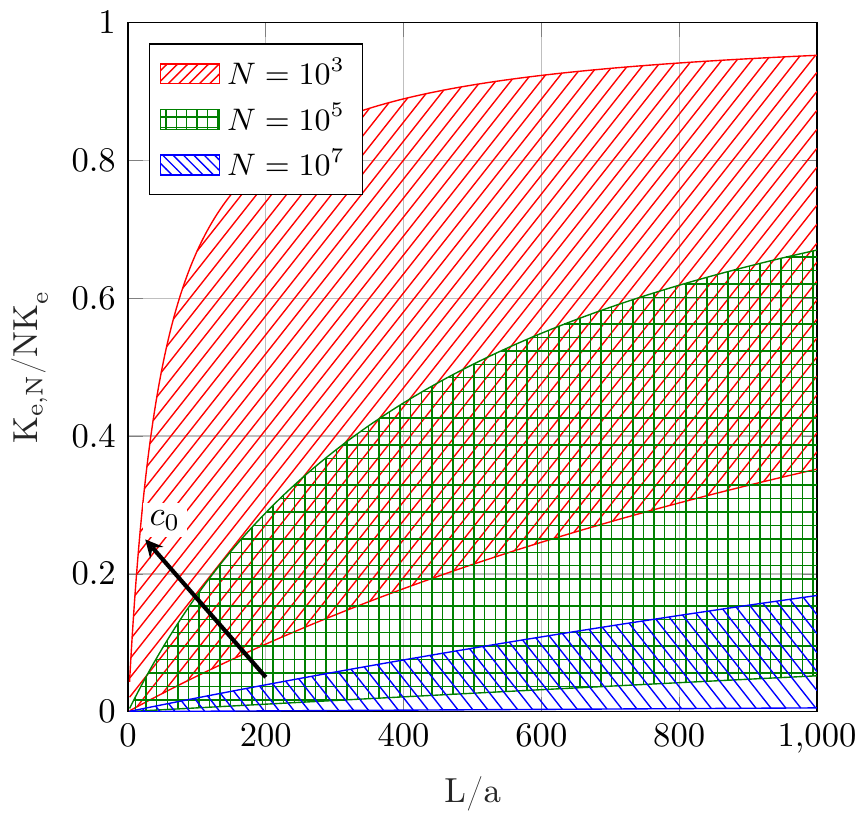}
    \caption{Ratio between total electric conductance $K_{e,N}$ and the one for N isolated pore $K_{e,N}^{isolated}=NK_e$ versus $L/a$ for a square-lattice membrane. Three $N$ are plotted. {For each of them, a hatched region is plotted to represent the region where the ratio spreads with the salt concentration in }the range $l_{Du}/a\in[2.6\times10^{-3},26]$ corresponding to salt concentration $c_0\in[1,10^{-4}]\,$M. {The black arrow shows the evolution of $c_0$ between each pair of boundary curves.} For the computation, we used $|\Sigma|=20$\,mC.m$^{-2}$, $h=50\,$nm and $a=25\,$nm.}
    \label{KeNsurNKe_square}
\end{figure}

We consider here again a square-lattice membrane. Fig.~\ref{KeNsurNKe_square} shows $\frac{K_{e,N}}{NK_{e}}$ as a function of $L/a$ ratio for {various } values of $N$. To compute this ratio for an aqueous solution of potassium chloride, we fixed  $|\Sigma|=20$\,mC.m$^{-2}$ (consistent for Si$_3$N$_4$ membranes \cite{lee2012}), $a=25\,$nm and $h=50\,$nm. The ratio is plotted in a range $l_{Du}/a\in[2.6\times10^{-3},26]$ corresponding to salt concentration $c_0\in[1,10^{-4}]\,$M. {To visualize the region where $\frac{K_{e,N}}{NK_{e}}$ spreads both with $L/a$ and the salt concentration, a hatched area is drawn for each $N$. The black arrow, common to each $N$, represents the way the salt concentration evolves between the bottom and the top curve (for each $N$)}.  {Note that $c_0=10^{-4}\,$M corresponds to $\kappa^{-1}=30\,$nm, which is a limit for non-overlapping EDL inside the pore.} 

Contrary to hydraulic permeability, evolution of $\frac{K_{e,N}}{NK_{e}}$ spreads on a large $L/a$ range. Nevertheless, the ratio tends asymptotically to 1, which is consistent to more and more isolated pores (and so to decreasing pore interactions). The normalized electric conductance is always lower than one, revealing a deleterious effect of pore interaction on electric transport efficiency, as already observed experimentally \cite{gadaleta2014} {in few-pore devices}. This effect decreases when $L/a$ rises but remain important even for very loosely porous membranes. Furthermore, we observe a strong influence of the number of pores on electric conductance. Contrary to hydraulic permeability, pore interactions are long-range since they scale as $(L/a)^{-1}$ whose infinite sum does not converge. Consequently, even if the influence of a distant pore is lower than for a neighbour one, the crescent number of interacting pores counterbalances {the decrease of the interaction strength, for a fixed $L/a$}. Finally, influence of the Dukhin length is noticeable. For a given $N$, the electric transport is larger when $l_{Du}$ is short (corresponding to high salt concentration), with ratio up to $\sim 10$ at low $L/a$.

Similarly to hydraulic permeability, we do not observe influence of membrane spatial organisation on electrical transport, when plotted as a function of membrane porosity (not shown here).



\subsection{Coupled transport}
\begin{figure}[h!]
    \centering
    \includegraphics[width=0.95\linewidth]{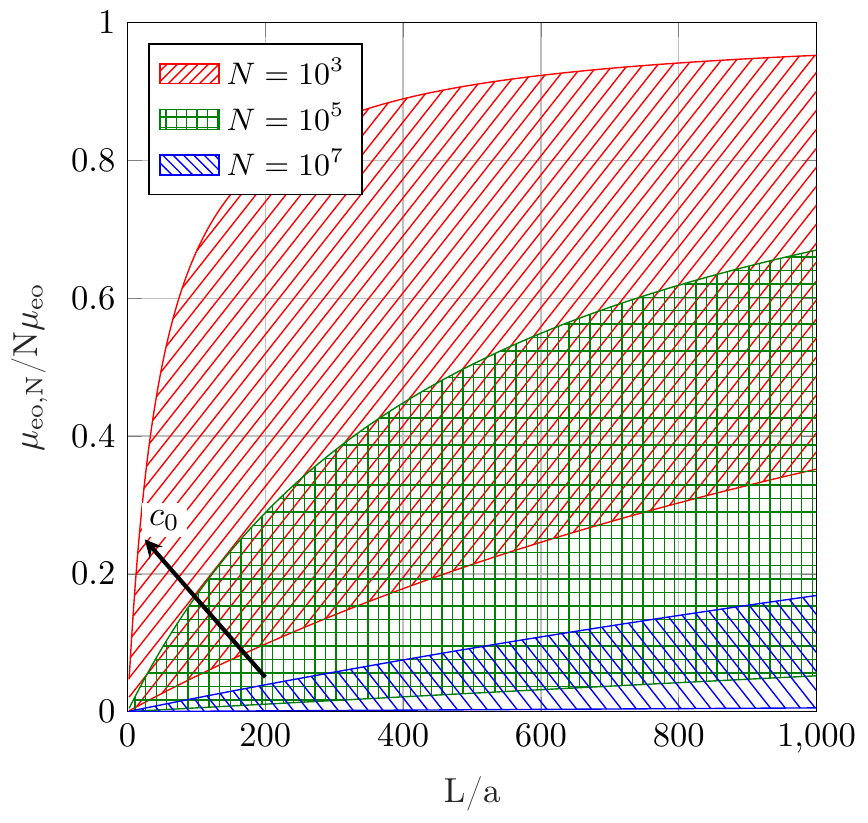}
    \caption{Ratio between total electroosmotic mobility $\mu_{eo,N}$ and the one for N isolated pore $\mu_{eo,N}^{isolated}=N\mu_{eo}$ versus $L/a$ for a {square}-lattice membrane. Three $N$ are plotted. {For each of them, a hatched region is plotted to represent the region where the ratio spreads with the salt concentration in }the range $l_{Du}/a\in[2.6\times10^{-3},26]$ corresponding to salt concentration $c_0\in[1,10^{-4}]\,$M. {The black arrow shows the evolution of $c_0$ between each pair of boundary curves.} For the computation, we used $|\Sigma|=20$\,mC.m$^{-2}$, $h=50\,$nm and $a=25\,$nm.}
    \label{mueoNsurNmueo_hexa}
\end{figure}

We {still} consider a {square}-lattice membrane. Fig. \ref{mueoNsurNmueo_hexa} shows $\mu_{eo,N} / N\mu_{eo}$ as a function of $L/a$ ratio for different $N$. To compute this ratio, we use the same conditions as the ones used for fig.~\ref{KeNsurNKe_square}.

The evolution of the electroosmotic mobility is very close to the one of the electric conductance. The evolution of $\frac{\mu_{eo,N}}{N\mu_{eo}}$ spreads on a large $L/a$ range but the ratio tends asymptotically to 1. The electroosmotic mobility is always lower than one, revealing a deleterious effect of pore interaction on coupled transport efficiency. This effect decreases when $L/a$ rises but remains important even for very loosely porous membranes. Furthermore, we observe a strong influence of the number of pores on coupled transport, similarly to electric conductance. Finally, the influence of the Dukhin length is also noticeable. For a given $N$, the electric transport is larger when $l_{Du}$ is short (corresponding to high salt concentration), with ratio up to~$\sim 10$ at low $L/a$. 

Such similar behaviour between $\mu_{eo,N}$ and $K_{e,N}$ can be explained from the expression obtained earlier for the electroosmotic mobility. We can observe that $\beta_{eo}^{(m)}=\beta_e^{(m)}\beta_h^{(m)}$. Yet, we observed that $K_{h,N}/NK_h$ tends to one for $L/a\sim 10$ so $\beta_h^{(m)}$ is asymptotically constant for $L/a\gg 1$. Thus, for large $L/a$, $\mu_{eo,N}\propto K_{e,N}$. The electroosmotic transport (or streaming current) in a multi-pore membrane is thus largely driven by the electrical transport, {and not by the hydrodynamic one}, when the inter-pore distance is larger than few pore radii.

\begin{figure}[h!]
    \centering
    \includegraphics[width=0.95\linewidth]{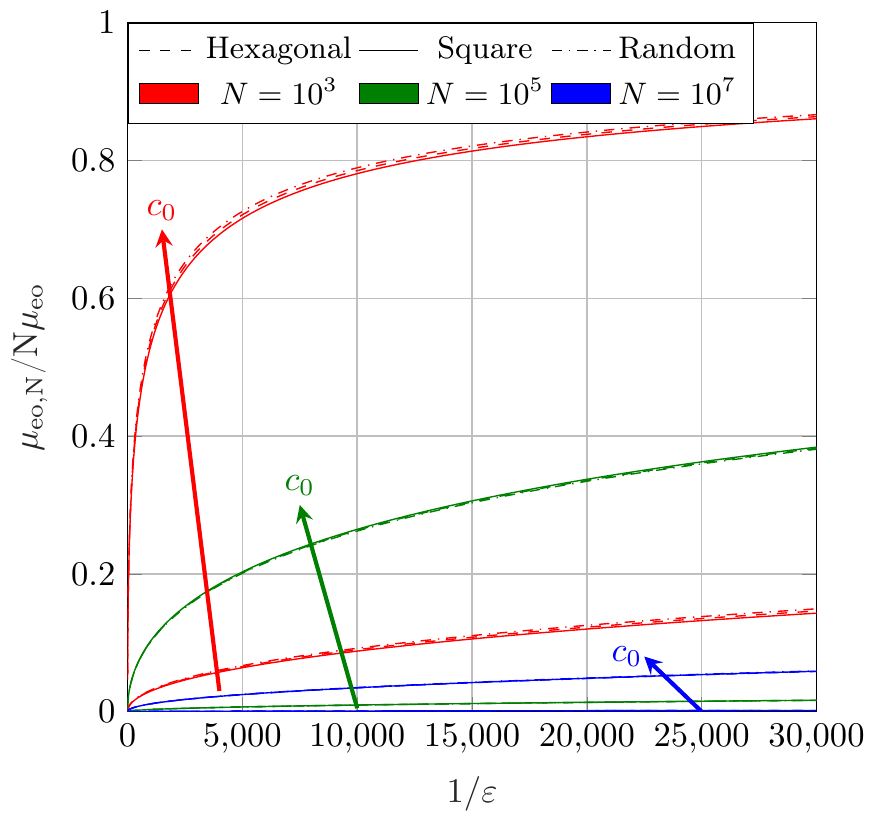}
    \caption{Ratio between total electroosmotic mobility $\mu_{eo,N}$ and the one for N isolated pore $\mu_{eo,N}^{isolated}=N\mu_{eo}$ versus $1/\varepsilon$. Square lattice, hexagonal lattice and random lattice are compared. Three $N$ are plotted. For each couple color/line style, two curves are presented. The top one corresponds to $l_{Du}/a=2.6\times10^{-3}$ ($c_0=1\,$M) and the bottom one corresponds to $l_{Du}/a=26$ ($c_0=10^{-4}\,$M). {Each arrow indicates the evolution direction of $c_0$ on the graph for a given $N$.} For the computation, we used $|\Sigma|=20$\,mC.m$^{-2}$, $h=50\,$nm and $a=25\,$nm.}
    \label{mueoNsurNmueo_comparaison}
\end{figure}

Fig.~\ref{mueoNsurNmueo_comparaison} shows a comparison of $\mu_{eo,N}/ N\mu_{eo}$ for three different membrane configurations: hexagonal lattice, square lattice and random lattice, as a function of the membrane porosity. We do not observe any noticeable differences when changing the pore spatial organisation. The porosity is {again} the only relevant parameter for coupled transport efficiency. 

\begin{figure}[h!]
    \centering
    \includegraphics[width=0.95\linewidth]{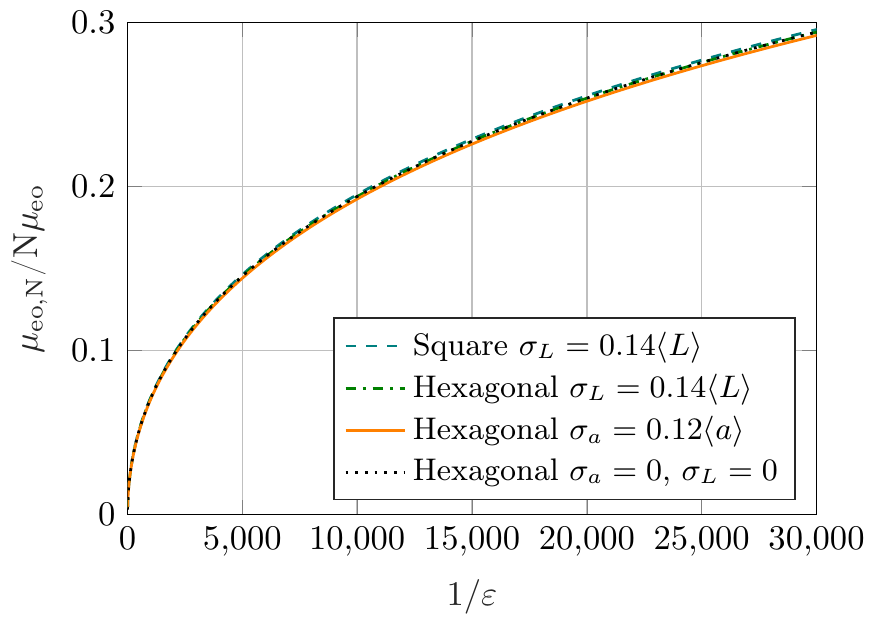}
    \caption{{Ratio between total electroosmotic mobility $\mu_{eo,N}$ and the one for $N=10^5$ isolated pore $\mu_{eo,N}^{isolated}=N\mu_{eo}$ versus $1/\varepsilon$ for disordered hexagonal and square lattice ($L$ follows an uniform distribution of width $\sigma_L=0.14\langle L\rangle$ where $\langle L\rangle$ is the mean inter-pore distance); and for a hexagonal lattice with variable pore size ($a$ follows an uniform distribution of width $\sigma_L=0.12\langle a\rangle$ where $\langle a\rangle$ is the mean pore radius, with conservation of the porosity). $l_{Du}/a=2.6\times10^{-1}$ ($c_0=10^{-2}\,$M), $|\Sigma|=20$\,mC.m$^{-2}$, $h=50\,$nm and $a=25\,$nm.} {For comparison, the case of a hexagonal lattice without noise is also plotted.}}
    \label{mueonoise}
\end{figure}

{We also consider the case of membrane with an initial square or hexagonal lattice who was \textit{noised} with different uniform distribution widths of distance between the pores (pseudo random membranes), and we once again observed a collapse when plotted as a function of the membrane porosity (Fig.~\ref{mueonoise}).}
To strengthen that {$\epsilon$} is indeed the only relevant parameter to consider, we investigated {the effect of the pore size, by considering now a membrane with a distribution of pore size around a mean value (but constrained by porosity conservation). In this case again, as shown in Fig.~\ref{mueonoise}, no deviation is observed, indicating that $\epsilon$ is indeed the relevant parameter to consider for membrane fabrication dedicated to EK effects}. {Once again, the variations observed by Jensen {et al.} \cite{jensen2014} in hydraulic resistance when they change pore radius distribution are uniquely due to porosity changes. When enlarging pore radius distribution symmetrically around a mean value, the pore surface distribution, which scales as $\langle a \rangle^2$, reveals a skewness towards high pore-surface values. Consequently, membrane porosity is enhanced, leading to reduced hydraulic resistance.}



\begin{figure}[h!]
    \centering
    \includegraphics[width=0.95\linewidth]{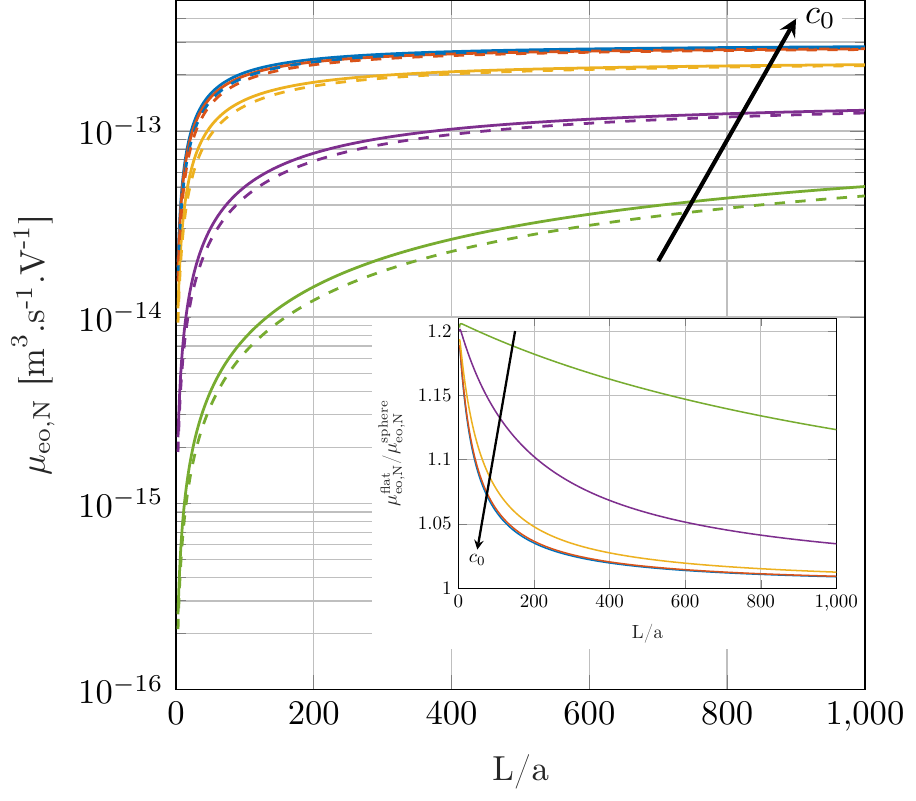}
    \caption{Electroosmotic mobility $\mu_{eo,N}$ versus $L/a$ for $N=10^3$ hexagonal-lattice flat and spherical membranes. Solid lines and dashed lines are related to a flat membrane and a spherical membrane respectively. Salt concentrations are $c_0\in[10^{-4};10^{-3};10^{-2};10^{-1},1]\,$M. Each color corresponds to one concentration. For the computation, we used $|\Sigma|=20$\,mC.m$^{-2}$, $h=50\,$nm and $a=25\,$nm. Inset: ratio between electroosmotic mobility $\mu_{eo,N}$ for a flat and a spherical membrane as a function of $L/a$ for $N=10^3$ pores.}
    \label{comparaison_flat_sphere}
\end{figure}

{The spherical membrane we have considered allows to reduce drastically computation time when $N$ becomes large{, which is a good way to explore a large range of number of pores, and different configurations}. Only one computation of the $\beta^{(m)}$ coefficients is {indeed} necessary, instead of one for each pore. If the study of pores placed on a sphere is justified with potential applications, and natural objects, flat membranes are also common in industry. {We seek now to estimate the error we get when considering a spherical membrane compared to a flat one. For this purpose} we performed the computation for a flat membrane pierced with $N=10^3$ pores. The values of the different Onsager's coefficients for a flat membrane do not exceed 20\% more than the ones for a spherical membrane, and the global trends with $c_0$ and $L/a$ remain similar. Fig. \ref{comparaison_flat_sphere} compares {the electroosmotic mobility in} these two situations {(spherical and flat membranes),} for different salt concentrations. One observes that the gap between a flat and a spherical membrane decreases {when $L/a$ rises}. Moreover, the difference drops faster when salt concentration is increased. This computation justifies to extend trends from spherical membranes to flat membranes {and confirms the use of spherical membranes to save computation time}.}

\subsection{Energy conversion efficiency}

\begin{figure}[h!]
    \centering
    \includegraphics[width=0.7\linewidth]{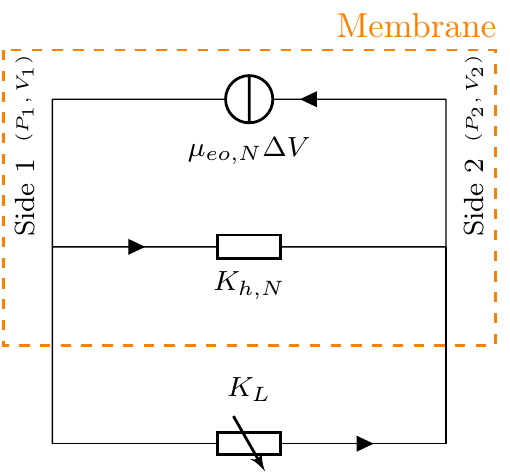}
    \caption{Sketch of the equivalent hydraulic circuit of a membrane connected to a load pipe.}
    \label{schema_circuit}
\end{figure}

The performance of a membrane for renewable energy recovery is evaluated by determining the theoretical maximum power yield of the electrokinetic process \cite{van_der_heyden2006,van_der_heyden2007}. In the framework presented here, hydraulic energy can be converted to electricity (streaming current) or electric energy can be converted to mechanical fluid displacement (electroosmosis). Let us consider the second case. A membrane, of hydraulic permeability $K_{h,N},$ is connected to a variable load pipe (analogous to a variable load resistance) whose hydraulic permeability is noted $K_L$. A voltage difference is applied through the membrane which leads to a flow rate $Q=\mu_{eo,N}\Delta V$. Fig. \ref{schema_circuit} represents the equivalent hydraulic circuit, {where membrane properties, and in particular pore interactions, are hidden in the encircled region noted membrane. In this region, the electrical and hydraulic analog of the membrane is depicted in fig.\ref{scheme}, bottom}. We follow the same reasoning as van der Heyden {et al.} \cite{van_der_heyden2006} for streaming current recovery through a nanochannel. An electric analogy allows to write:

\begin{equation}
    \Delta P=-\frac{\mu_{eo,N}\Delta V}{K_{h,N}+K_L}.
\end{equation}

The injected (electric) power is $\mathcal{P}_{in}=I\Delta V$ whereas the recovered (mechanical) power can be written as $\mathcal{P}_{out}=K_L\Delta P^2$. Electric current $I$ can be expressed as $I=\mu_{eo,N}\Delta P+K_{e,N}\Delta V$. Consequently, efficiency of energy conversion is defined as: 

\begin{equation}
    \mathcal{E}=\frac{\mathcal{P}_{out}}{\mathcal{P}_{in}}=\frac{\mu_{eo,N}^2K_L}{K_{e,N}(K_{h,N}+K_L)^2-\mu_{eo,N}^2(K_{h,N}+K_L)}.
\end{equation}

\noindent We write out $\alpha=\mu_{eo,N}^2/(K_{h,N}K_{e,N})$ and $\Theta=K_{h,N}/K_L$. The efficiency of energy conversion can be written as: 

\begin{equation}
    \mathcal{E}=\frac{\alpha\Theta}{(1+\Theta)(1+\Theta-\alpha\Theta)}.
\end{equation}

\noindent The maximal efficiency is obtained for $\Theta=1/\sqrt{1-\alpha}$ (impedance matching). This leads to a maximal efficiency: 

\begin{equation}
    \mathcal{E}_{max}=\frac{\alpha}{\alpha+2\left(\sqrt{1-\alpha}+1-\alpha\right)}.
\end{equation}

\begin{figure}[h!]
    \centering
    \includegraphics[width=0.95\linewidth]{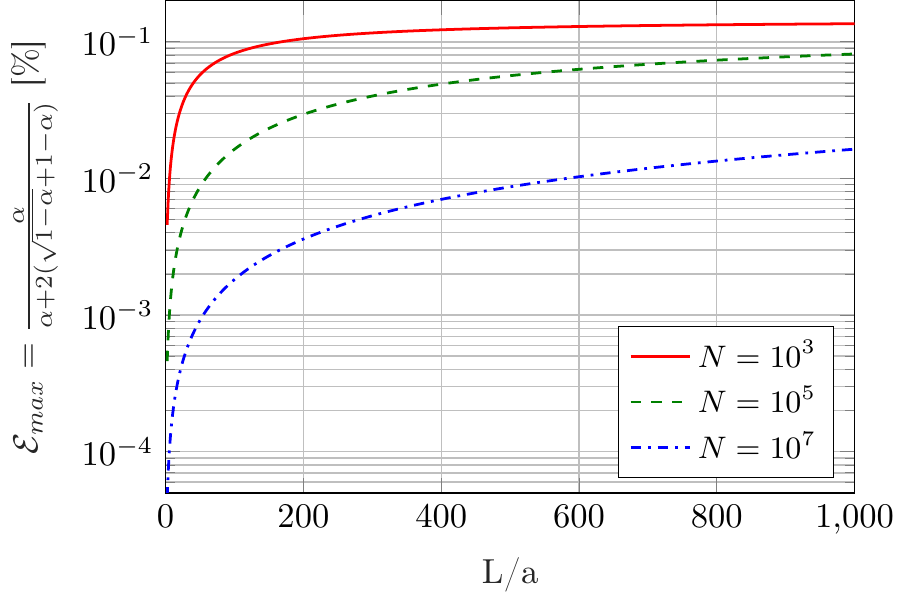}
    \caption{Maximal efficiency for electroosmotic transport through a hexagonal-lattice membrane for different $N$. $l_{Du}/a=2.6\times10^{-1}$ corresponding to salt concentration $c_0=10^{-2}\,$M.}
    \label{efficacite}
\end{figure}

Fig.~\ref{efficacite} shows the maximal yield $\mathcal{E}_{max}$ as a function of $L/a$, in a semi-log scale, for a hexagonal-lattice membrane with same parameters as in fig.~\ref{mueoNsurNmueo_hexa}. Numerical results are plotted for only one concentration, $c_0=10^{-2}\,$M ($l_{Du}/a=0.26$). {Results from previous sections show that there is no influence of the pore spatial organisation and that the main relevant parameter is again the porosity of the membrane $\varepsilon$. One notices however that efficiency  increases with $L/a$, meaning that pore interactions decrease energy conversion yield. Consequently, addition of pores is deleterious for energy conversion efficiency.}

\begin{figure}[h!]
    \centering
    \includegraphics[width=0.95\linewidth]{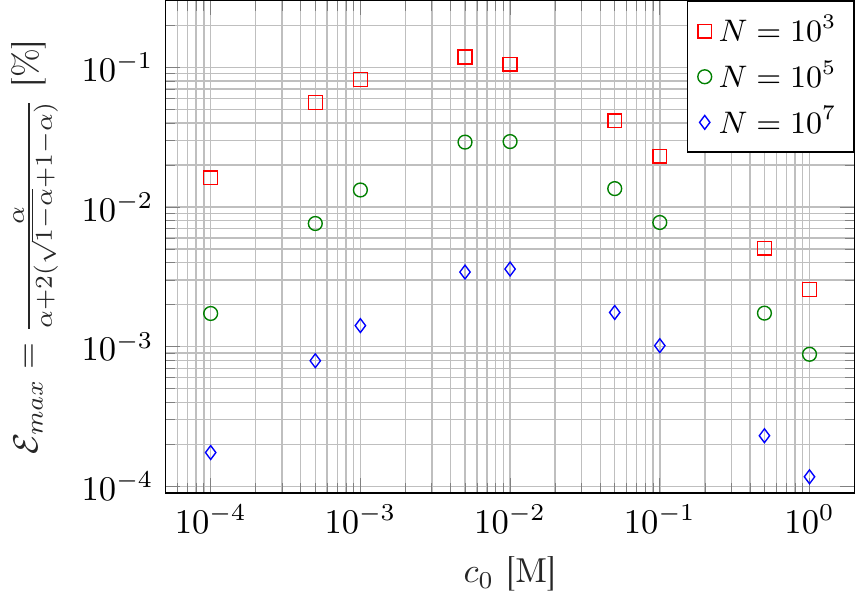}
    \caption{Maximal efficiency for electroosmotic transport through a hexagonal-lattice membrane for different $N$ versus salt concentration. $L/a$ is fixed to 200.}
    \label{efficacite_vs_c}
\end{figure}

A central phenomenon is the influence of the salt concentration. {The effect of salt concentration shows up in surface conduction term, characterized in eq. \ref{eq:kae} by the term depending on the Dukhin length $l_{Du}$. Indeed, as surface entrance effects have been shown to be important for the electric conductivity of one pore \cite{lee2012}, we also considered them for pore interactions (eq. \ref{eq:accessKam}, figs. \ref{KeNsurNKe_square} and \ref{mueoNsurNmueo_hexa})}. Fig.\ref{efficacite_vs_c} shows the maximal efficiency as a function of salt concentration for $L/a=200$. This efficiency is not monotonic, with a maximum around $5\times10^{-3}\,$M. The values computed here reach asymptotically 0.1\% for $c_0\sim10^{-2}\,$M and $N=10^3$, which seems to be near a maximal value in the case where pore interactions become negligible. Let's note that such a {dependency of the yield versus salt concentration} was observed experimentally \cite{van_der_heyden2007} for a {single} nanochannel, {but of course in this case with a different origin}. It is due to variations of surface charge density with respect to salt concentration \cite{van_der_heyden2007}. 
{A thorough study should take into account both contributions}.



\section{Discussion and conclusion}

{We have investigated electrokinetic coupled transport properties through a nanofluidic membrane drilled with a large number of nanopores, by taking into account entrance effects and then pair interactions between the pores. This study leads to various results, which are crucial when the design of membrane for renewable energy conversion  is considered.}

First, {using orders of magnitude and already documented scalings, we have estimated} that the electroosmotic mobility scales as $\sqrt{N}$ {for large $N$,} where $N$ is the number of pore in the membrane. This sub-linear scaling is the same as the one obtained  for electric transport in the membrane \cite{gadaleta2014}, the effect of interactions between pores in the hydraulic resistance being incidentally very small compared to the electrical case \cite{jensen2014}.

{Second, we have shown for the first time that{, for a given membrane porosity,} the different transport coefficients do not depend on the spatial organisation of the pores. When normalized by membrane porosity, all transport coefficients collapse whatever the membrane type (amorphous arrangement, hexagonal-lattice, square-lattice) and the pore size distribution.}  A possible interpretation for this absence of effect is that the high number of pores ($N\geq10^3$) smooth the spatial organisation, provided the porosity is the same.
{This is a crucial result in the design of nanoporous membrane.} 
Furthermore, whereas pore interactions are beneficial for hydrodynamic transport, up to $L/a\sim 10$, they are deleterious for electric transport, even for $L/a\gg10$.  This is due to the $(L_{n,m}/a)^{-1}$ dependency ($L_{n,m}$ is the distance between pore $m$ and pore $n$) of the electric conductance of a given pore $m$ in the membrane. The term $\gamma^{(m)}$ looks like a harmonic series which does not converge. In our study, since the sum is finite, there is no convergence problem, but the variation is slower than the one of $\lambda^{(m)}$ which is at maximum a sum of $(L_{n,m}/a)^{-3}$ terms. Since we can write, for a spherical membrane where all pores have the same environment:
\begin{equation}
    \mu_{eo,N} = \frac{K_{h,N}K_{e,N}}{K_h^iK_e^i}\frac{\mu_{eo}^i}{N},
\end{equation}
\noindent the behaviour of $\mu_{eo,N}$ is the same as $K_{e,N}$ for $L/a\gtrsim 10$ ({$\mu_{eo,N} \simeq \mu_{eo}^i K_{e,N} / K_{e,i}$}).

{These opposite behaviours can be interpreted physically. The hydrodynamic streamlines focusing at the pore entrance lead to energy dissipation, and so hydraulic resistance increase. When several pores interact, this tightening is pooled and it leads to a decrease of the global hydraulic resistance compared to isolated pores. Concerning the access ionic conductance, it can be determined by making an electrostatic analogy \citep{gadaleta2014}, each pore entrance can be considered as a disk holding an electric charge. One can compute a capacitance $C$ related to this charge $q$ and the electric potential $V_0$ at the pore entrance ($q=CV_0$). The access resistance is by definition inversely proportional to the capacitance ($R_e^a=\varepsilon_0\varepsilon_r/(\kappa_b C)$). The mutual pore interactions should lead to an electric potential increase. Since we work at fixed potential, the charge held by each disk must decrease, which means a decrease of the subsequent capacitance, and so an increase of the access resistance of each pore.}







    

Third, we show that both the electric conductance and the electroosmotic mobility varies a lot with salt concentration. This can be explained by the origin of pore interactions. They are due to access resistance provoked, among others, by surface conduction effects \cite{lee2012}. {They are directly related to the extension of the zone at which surface conduction effects dominate the bulk ones, i.e. the Dukhin length $l_{Du} \sim |\Sigma|/c_0$.}
A low salt concentration results in a large Dukhin length and enhanced entrance effects. Since pore interactions are deleterious for electric and electroosmotic transport, a drop in salt concentration results in {a} decrease of {the} related Onsager's coefficients.

{Fourth, we calculate the energy conversion efficiency of such nanofluidic membranes and we show} it reveals an exotic behaviour. {For a fixed surface charge density,} it is not monotonic with salt concentration, with a maximum around $c_0=5\times10^{-3}$\,M, whereas $K_{e,N}$ and $\mu_{eo,N}$ are monotonic with $c_0$ for a fixed $L/a$ ratio (as reported in Fig.~\ref{comparaison_flat_sphere}). {Theoretically, a plateau is expected for efficiency as a function of concentration \cite{van_der_heyden2006}. It is due to surface-conduction-generated plateau in electric conductance versus salt concentration \cite{schoch2008}. This plateau of efficiency is observed experimentally in a nanochannel \cite{van_der_heyden2007} with a subtle peak when EDL overlap. In our case, we do not consider EDL overlapping and minimal concentration was chosen to avoid it. So we should expect a plateau. However, we can attribute the peak observed on $\mathcal{E}_{max}$ (fig. \ref{efficacite_vs_c}) to surface effects influence on pore interactions. In addition to the electric resistance of the pores, two phenomena are in competition to explain energy conversion efficiency. (i) Surface effects, related to $l_{Du}$, are deleterious for pore interactions and decrease the coupled transport amplitude (fig. \ref{comparaison_flat_sphere}). An increase of $c_0$ (decrease of $l_{Du}$) is good for coupled transports. (ii) Presence of co-ions in the pore, out of EDL, participates to power dissipation, but not to coupled transports \cite{van_der_heyden2006}. Consequently, an increase of salt concentration, {via} the EDL thickness decrease, reduces $\mathcal{E}_{max}$. The first effect is more important when $a\ll l_{Du}$ (low $c_0$) whereas the second one is important when $a\gg \kappa^{-1}$ (high $c_0)$. This explains the appearance of such a peak of $\mathcal{E}_{max}$ in the intermediate zone of salt concentration. }


{This work constitutes a first step to optimize the design of nanoporous membranes for energy harvesting. A trade-off between low porosity to reach larger yields and space constraints to add as many channels as possible in a membrane has to be reached. {If primary energy is ``free'' (osmotic pressure for instance)}, {recoverable power $\mathcal{P}_{out}$ is proportional to $N$ for a large number of pores, provided that the hydrodynamic load is large. It means that the recoverable power per unit of membrane surface varies as $1/N$ or, if the porosity is constant, as $1/\sqrt{S}$ if $S$ is the surface of the membrane. Larger membranes will be indeed more efficient than smaller ones, but the sublinear relationship shows that the quest to always larger membranes might not be the ultimate strategy. Specific design of small membrane patches can indeed be more efficient. Finally,}  a new effect of salt concentration due to access resistance has also to be considered and to be coupled to other concentration effects such as interface ionization. Future directions should concern (i) the experimental evidences of these predictions, (ii) the consideration of other types of EK transport or ways of nanofluidics energy harvesting (chemical or thermal energy harvesting for example) and (iii) the investigation of membranes with much more complex geometries, such as connected entangled channels.}

\bigskip 

{\section*{Acknowledgements}
The authors thank S. Gravelle, L. Bocquet, A. Parrenin and J.-D. Julien for fruitful discussions.
We would like to thank the ANR through the projects Blue Energy (ANR-14-CE05-0017) and NECtAR (ANR-16-CE06-0004-01) for funding. A.-L. Biance thanks IDEX of University Lyon 1 through the Elan-ERC program, and C. Sempere thanks financial support from \'Ecole Normale Sup\'erieure de Lyon.  }

\nomenclature{$\kappa^{-1}$}{Debye length}
\nomenclature{$I_s$}{Ionic strength}
\nomenclature{$c_i$}{Concentration of ionic specie $i$}
\nomenclature{$z_i$}{Valence of ionic specie $i$}
\nomenclature{$Q$}{Fluid flow rate}
\nomenclature{$I$}{Electric current}
\nomenclature{$K_h$}{Total hydrodynamic permeability of one pore}
\nomenclature{$K_h^i$}{Inner hydrodynamic permeability of one pore}
\nomenclature{$K_h^a$}{Access hydrodynamic permeability of one pore}
\nomenclature{$K_e$}{Total electric conductance of one pore}
\nomenclature{$K_e^i$}{Inner electric conductance of one pore}
\nomenclature{$K_e^a$}{Access electric conductance of one pore}
\nomenclature{$\Delta V$}{Total potential difference through the pore}
\nomenclature{$\Delta V^i$}{Inner potential difference through the pore}
\nomenclature{$\Delta V^a$}{Access potential difference through the pore}
\nomenclature{$\Delta P$}{Total pressure drop}
\nomenclature{$\Delta P^i$}{Inner pressure drop}
\nomenclature{$\Delta P^a$}{Access pressure drop}
\nomenclature{$c_i$}{Concentration of ionic specie $i$}
\nomenclature{$\mu_{eo}$}{Total electroosmotic mobility of one pore}
\nomenclature{$\mu_{eo}^i$}{Inner electroosmotic mobility of one pore}
\nomenclature{$\mu_{eo,N}$}{Electroosmotic mobility of a membrane of $N$ pores}
\nomenclature{$k_h$}{$K_h^i/K_h^a$}
\nomenclature{$k_e$}{$K_e^i/K_e^a$}
\nomenclature{$\alpha^i$}{$(\mu_{eo}^i)^2/(K_h^iK_e^i)$}
\nomenclature{$\alpha$}{$(\mu_{eo,N})^2/(K_{h,N}K_{e,N})$}
\nomenclature{$\Theta$}{$K_{h,N}/K_L$}
\nomenclature{$h$}{Pore length}
\nomenclature{$a$}{Pore radius}
\nomenclature{$K_{e,N}^i$}{Inner electric conductance of a membrane of $N$ pores}
\nomenclature{$K_{h,N}^i$}{Inner hydrodynamic permeability of a membrane of $N$ pores}
\nomenclature{$R_h^a$}{Access electric resistance of one pore}
\nomenclature{$\eta$}{Dynamic viscosity of the fluid}
\nomenclature{$R_h^i$}{Inner electric resistance of one pore}
\nomenclature{$R_h^{a,(m)}$}{Access electric resistance of the pore labelled $m$}
\nomenclature{$L_{n,m}$}{Center-to-center distance between pores $n$ and $m$}
\nomenclature{$L$}{Typical center-to-center distance between adjacent pores}
\nomenclature{$\lambda^{(m)}$}{Geometric factor for hydrodynamic resistance}
\nomenclature{$\Lambda$}{Asymptotic value of $\lambda^{(m)}$}
\nomenclature{$N$}{Total number of pores of the membrane}
\nomenclature{$l_{Du}$}{Dukhin length}
\nomenclature{$\varepsilon_0\varepsilon_r$}{Dielectric constant}
\nomenclature{$\zeta$}{Zeta potential}
\nomenclature{$c_0$}{Salt concentration}
\nomenclature{$\kappa_b$}{Bulk electric conductivity}
\nomenclature{$\kappa_s$}{Surface electric conductivity}
\nomenclature{$\lambda_{K^+}$}{Potassium ionic molar conductivity}
\nomenclature{$\lambda_{Cl^-}$}{Chlorine ionic molar conductivity}
\nomenclature{$\beta_h$}{$1/(1+2k_h)$}
\nomenclature{$\beta_e$}{$1/(1+2k_e)$}
\nomenclature{$\beta_{eo}$}{$1/[(1+2k_h)(1+2k_e)]$}
\nomenclature{$\beta_h^{(m)}$}{$1/(1+2k_h^{(m)})$}
\nomenclature{$\beta_e^{(m)}$}{$1/(1+2k_e^{(m)})$}
\nomenclature{$\beta_{eo}^{(m)}$}{$1/[(1+2k_h^{(m)})(1+2k_e^{(m)})]$}
\nomenclature{$K_h^{(m)}$}{Hydrodynamic permeability of the pore labelled $m$}
\nomenclature{$K_h^{a,(m)}$}{Access hydrodynamic permeability of the pore labelled $m$}
\nomenclature{$K_e^{(m)}$}{Electric conductance of the pore
labelled $m$}
\nomenclature{$K_e^{a,(m)}$}{Access electric conductance of the pore labelled $m$}
\nomenclature{$\mu_{eo}^{(m)}$}{Electroosmotic mobility of the pore labelled $m$}
\nomenclature{$a^{eff}$}{$a+0.5l_{Du}$}
\nomenclature{$\gamma^{(m)}$}{Geometric factor for electric conductance}
\nomenclature{$K_{h,N}^{isolated}$}{Hydrodynamic permeability if $N$ pores are isolated}
\nomenclature{$K_{e,N}^{isolated}$}{Electric conductance if $N$ pores are isolated}
\nomenclature{$\mu_{eo,N}^{isolated}$}{Electroosmotic mobility of a membrane with $N$ isolated pores}
\nomenclature{$\varepsilon$}{Membrane porosity}
\nomenclature{$S_{membrane}$}{Membrane area}
\nomenclature{$a_m$}{Radius of the pore labelled $m$}
\nomenclature{$K_L$}{Load pipe's hydrodynamic permeability}
\nomenclature{$\mathcal{P}_{in}$}{Injected (electric) power}
\nomenclature{$\mathcal{P}_{out}$}{Recovered (mechanical) power}
\nomenclature{$\mathcal{E}$}{Efficiency of energy conversion}
\nomenclature{$\mathcal{E}_{max}$}{Maximal energy conversion efficiency}
{\printnomenclature}

\bibliography{biblio_ILM}
\end{document}